\documentclass[iop]{emulateapj}

\usepackage{times,color,natbib,url,amsmath}
\usepackage{graphicx,float,psfrag}
\usepackage{tabularx}
\usepackage{latexsym,amsmath,amssymb}
\usepackage{natbib}
\usepackage{verbatim}
\usepackage{multirow}

\bibpunct{(}{)}{;}{a}{}{,}

\newcommand {\asca} {{\it ASCA}}

\newcommand {\sax} {BeppoSAX}

\newcommand {\xmm} {\textsl{XMM-Newton}}
\newcommand {\chandra} {\textsl{Chandra}}

\newcommand {\rxte} {\textsl{RXTE}}

\def \rsun {\ifmmode$R$_{\odot}\else R$_{\odot}$}

\def \hcm {\hbox {\ifmmode $ atoms cm$^{-2}\else atoms cm$^{-2}$\fi}}

\def\approxgt{\mathrel{\hbox{\rlap{\lower.55ex \hbox {$\sim$}}
        \kern-.3em \raise.4ex \hbox{$>$}}}}
\def\approxlt{\mathrel{\hbox{\rlap{\lower.55ex \hbox {$\sim$}}
        \kern-.3em \raise.4ex \hbox{$<$}}}}

\def \src {SGR\,1806$-$20}

\begin{document}

\title{\xmm\ observations of \src\ over seven years following the 2004 Giant Flare}

%
%

\author{G.~Younes$^{1}$, C.~Kouveliotou$^{1}$, V.~M.~Kaspi$^{2}$}

 \affil{
$^1$ Department of Physics, The George Washington University, Washington, DC 20052, USA \\
$^2$ Department of Physics, McGill University, Montreal, Quebec, H3A 2T8, Canada \\
}

%
%

\begin{abstract}

We report on the study of 14 \xmm\ observations of the magnetar \src\
spread over a period of 8 years, starting in 2003 and extending to
2011. We find that in mid 2005, a year and a half after a giant
flare (GF), the torques on the star increased to the largest value yet
seen, with a long term average rate between 2005 and 2011 of
$\lvert\dot{\nu}\rvert\approx1.35\times10^{-11}$~Hz~s$^{-1}$, an order 
of magnitude larger than its historical level measured in 1995. The
pulse morphology of the source is complex in the observations
following the GF, while its pulsed-fraction remained constant at about
$7\%$ in all observations. Spectrally, the combination of a black-body
(BB) and power-law (PL) components is an excellent fit to all
observations. The BB and PL fluxes increased by a factor of 2.5 and 4,
respectively, while the spectra hardened, in concordance with the 2004
major outburst that preceded the GF. The fluxes decayed exponentially
back to quiescence with a characteristic time-scale of
$\tau\sim1.5$~yrs, although they did not reach a constant value until
at least 3.5 years later (2009). The long-term timing and
  spectral behavior of the source point to a decoupling between the
  mechanisms responsible for their respective behavior. We argue that
low level seismic activity causing small twists in the open
field lines can explain the long lasting large torques on the star,
while the spectral  behavior is due to a twist imparted onto closed
field lines after the 2004 large outburst.

\end{abstract}

\section{Introduction}
\label{Intro}

Bursting activity in magnetars (neutron stars with X-ray emission
  powered by the decay of their strong internal magnetic fields) is
usually accompanied by changes in the spectral and temporal properties
of their persistent emission. During an active period, the X-ray flux
of the source increases and can occasionally reach up to three orders
of magnitude higher than its quiescence level \cite[see ][for a
review]{rea11:outburst}. This increase is usually accompanied by
spectral hardening \citep[][]{kaspi03ApJ:2259,mereghetti15:mag},
changes in the shape of the pulse profile and the pulsed fraction
\citep[e.g., ][]{muno07MNRAS:164710}, and variation in the source
timing properties, either in the form of a glitch
\citep{dib08ApJ:glitches,archibald13Natur} or a more gradual change of
the spin-down rate \citep{archibald15ApJ:1048}. A direct connection
between all these effects, however, is still not entirely clear
\citep[e.g.,][]{woods99ApJ:1900,woods02ApJ:1900,ng11ApJ:1547,dib14ApJ}. These
observations thus put tight constraints on any theoretical models
attempting to explain the X-ray emission mechanism of magnetars.

\src, one of the most prolific bursters and the last one to emit a
Giant Flare \citep[GF, December 2004,][]{hurley05NaturGF1806,
  gaensler05Natur:1806}, is a perfect example of a varying
magnetar. Its quiescent X-ray emission shows one of the most erratic
behaviors within the magnetar population, both in its timing and its
spectral properties. The source power-law spectrum hardened gradually
from an index of 2.2 in 1995 to 1.6 in 2003
\citep{mereghetti05ApJ:1806}. The pulse profile, on the other hand,
was consistently single pulsed during that time, and only changed to a
double-peaked profile after the GF \citep{woods07ApJ:1806}. The pulse
frequency and frequency derivative history of the source changed
dramatically in the last twenty years \citep[see Figure~\ref{freqHis}
of ][]{woods07ApJ:1806}. The latter decreased monotonically starting
in 1999, a few weeks after a small bursting activity episode, and
reached a value below its 1993 historical level. It flattened out in
2002 and remained constant up to 2004 when strong bursting activity
was recorded. However, a sudden and erratic change was seen with a
hysteresis relative to that activity lasting up to a few months after
the GF. In mid 2005, the period derivative increased back to the
pre-2004 bursting-episode level \citep{woods07ApJ:1806}.

Here, we revisit \src\ with emphasis on the long-term temporal and
spectral variations of the source persistent emission. We analyze all
publicly available \xmm\ observations that span eight years, from 2003
to 2011. Nine observations took place after the December 2004 GF
spreading over 7.5 years. The observations and data reduction are
summarized in Section~\ref{obs}; Section~\ref{res} shows the evolution
of the source temporal and spectral properties, which are then
discussed in Section~\ref{discuss}.

\section{Observations and data reduction}
\label{obs}

\src\ was observed with \xmm\ a total of 14 times over the span of 8
years, starting April 2003. Ten of these observations were performed
after the December 2004 GF. In all observations, the EPIC-pn
\citep{struder01aa} camera was operated in either Large Window (73-ms
resolution) or small window (6-ms resolution) mode, using the thin or
medium filter. All  data products were obtained from the \xmm\ Science
Archive (XSA)\footnote{http://xmm.esac.esa.int/xsa/index.shtml} and
reduced using the Science Analysis System (SAS) version 13.5.0. Data
are selected using event patterns 0--4, during only good X-ray events
(``FLAG$=$0'').  We applied the task {\sl
  epatplot}\footnote{http://xmm.esac.esa.int/sas/current/doc/epatplot/epatplot.html}
to all observations. This task allows for a pile-up estimate through
the direct comparison of the fraction of the event patterns in a given
event file to model curves from a calibration file, e.g.,
EPN\_QUANTUMEF\_0016.CCF for PN data. The pattern fraction followed the
model perfectly and we concluded that none were affected by
pile-up. Most observations, however, showed intervals of high
background. In such cases, we excluded time periods where the
background level was higher than 5\%\ of the source flux. We also
excluded time intervals associated with bursts from the source
\citep[e.g., ][]{mereghetti05ApJ:1806,esposito07AA:1806}. Finally, we
excluded the  MOS cameras from our analysis due to the poorer timing
resolution and collecting area. Only one observation had both MOS1 and
MOS2 operating in timing mode (1.75~ms resolution, 0164561101), 6 more
observations had only MOS1 operating in timing mode (0164561301,
0164561401, 0406600301, 0406600301, 0502170301, 0502170401), while the
rest were operated in either full-frame (2.6~s resolution, 0554600301,
0554600401, 0654230401) or large window (0.9~s resolution, 0148210101,
0148210401, 0205350101, 0604090201) modes. The log of the 14 \xmm\
observations we analyzed is listed in Table~\ref{logObs}.

\xmm\ EPIC-pn source events for all observations were extracted from a
circle with center and radius obtained by running the task  {\sl
  eregionanalyse}  on the cleaned event files. This task calculates
the optimum centroid of the counts distribution within a given source
region, and the radius of a circular extraction region that maximizes
the source signal to noise ratio. Background events are extracted from
a source--free circle with the same radius as the source on the same
CCD. We generated response matrix files using the SAS task {\sl
  rmfgen}, while ancillary response files were generated using the SAS
task {\sl arfgen}. The EPIC spectra were created in the energy range
0.5--10~keV, and grouped to have a signal  to noise ratio of 6 with a
minimum of 30 counts per bin to allow the use of the $\chi^2$
statistic.

\begin{table}[t]
\caption{Log of the \xmm\ observations along with their timing properties}
\label{logObs}
\newcommand\T{\rule{0pt}{2.6ex}}
\newcommand\B{\rule[-1.2ex]{0pt}{0pt}}
\begin{center}{
\resizebox{0.48\textwidth}{!}{
\begin{tabular}{c c c c c}
\hline
\hline
Observation ID   \T\B & Date & GTI Exposure & $\nu$ (error) & PF (error)$^{a}$ \\
                           \T\B &         &      (ks)     &    (Hz) &       \\
\hline
0148210101$^b$ \T & 2003-04-03 & 4.6 & 0.132784 (7)  & 0.05 (0.02)\\
0148210401$^b$ \T & 2003-10-07 & 9.7 & 0.1326236 (4)  & 0.07 (0.02)\\
0205350101$^b$ \T & 2004-09-06 & 43.0 & 0.1323468 (7)  & 0.08 (0.01)\\
0164561101$^b$ \T & 2004-10-06 & 17.9 & 0.132325 (8)  & 0.06 (0.01)\\
0164561301$^c$ \T & 2005-03-07 & 3.3 &        0.13228 (2)  & 0.03 (0.02)\\
0164561401$^d$ \T & 2005-10-04 & 31.8 &        0.132156 (1)  & 0.07 (0.01)\\
0406600301$^d$ \T & 2006-04-04 & 47.5 &       0.131909 (3)  & 0.06 (0.02)\\
0406600401$^e$ \T & 2006-09-10 & 30.8 & 0.131771 (3)  & 0.06 (0.03)\\
0502170301 \T & 2007-09-26 & 30.5 &        0.131430 (3)  & 0.07 (0.03)\\
0502170401 \T & 2008-04-02 & 31.0 &        0.131038 (3)  & 0.07 (0.02)\\
0554600301 \T & 2008-09-05 & 35.0 &        0.130869 (3)  & 0.07 (0.03)\\
0554600401 \T & 2009-03-04 & 29.6 &        0.130633 (2)  & 0.06 (0.02)\\
0604090201 \T & 2009-09-08 & 29.0 &        0.130441 (2)  & 0.06 (0.03)\\
0654230401 \T & 2011-03-23 & 30.0 &        0.129838 (1)  & 0.07 (0.02)\\
\hline
\end{tabular}}}
\end{center}
\begin{list}{}{}
\item[{\bf Notes.}]$^{a}$ RMS pulsed fraction in the
  1.5$-$10~keV energy range. Observations also analyzed in: $^{b}$
  \citet{mereghetti05ApJ:1806}, $^{c}$ \citet{tiengo05AA:1806}, $^{d}$
    \citet{mereghetti07ApSS:1806}, and $^{e}$ \citet{esposito07AA:1806}.
\end{list}
\end{table}

We performed our spectral analysis using XSPEC \citep{arnaud96conf}
version 12.8.1. The photo-electric cross-sections of
\citet{verner96ApJ:crossSect} and the abundances of \citet{wilms00ApJ}
are used throughout to account for absorption by neutral gas. All
quoted uncertainties are at the $1\sigma$ level, unless otherwise
noted.

\section{Results}
\label{res}

\subsection{Timing analysis}
\label{timeana}

\begin{figure}[]
\begin{center}
\includegraphics[angle=0,width=0.47\textwidth]{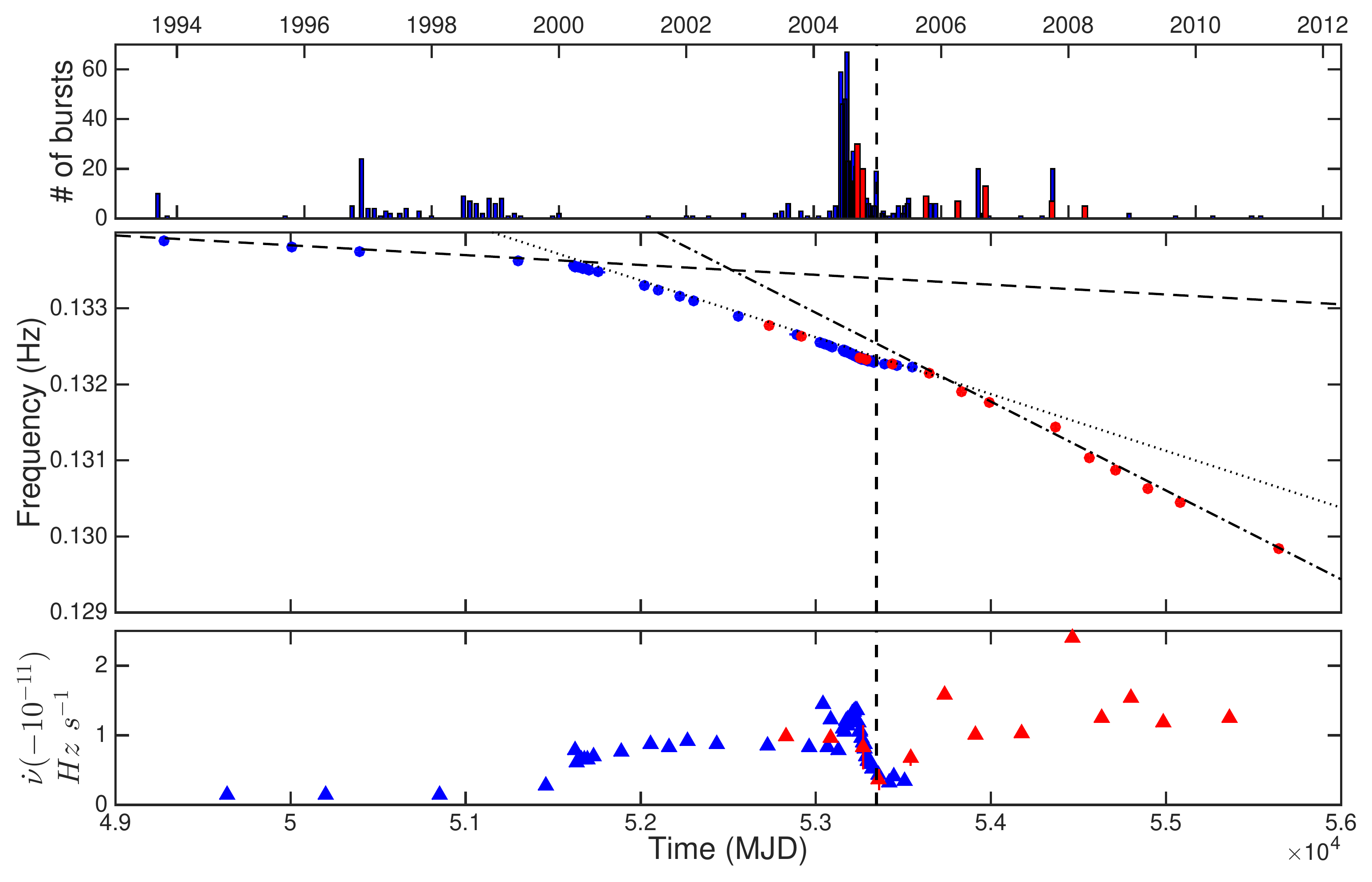}
\caption{Frequency and frequency derivative history of \src\ from mid
  1993 until mid 2011. {\sl Top panel.} Number of bursts (per 30-day
  intervals). The blue bars represent data collected from \citet{ 
    woods07ApJ:1806} up to 2005 June, and bursts reported in GCNs up
  to 2012. These represent bursts as seen with wide field-of-view
  instruments. We also included in red the bursts seen with \xmm. {\sl
    Middle panel.} Spin frequency history. Blue dots are collected
  from \citet{woods07ApJ:1806} and red dots are our \xmm\
  measurements. The dashed and dotted lines are fits to the frequency
  derivative from 1993 to 2000 January
  ($\dot{\nu}=-1.48\times10^{-12}$~Hz~s$^{ -1}$), and 2001 January to
  2004 April ($\dot{\nu}=-8.69\times10^{-12}$~Hz~s$^{ -1}$,
  \citealt{woods07ApJ:1806}). The dot-dashed line is our best fit to
  frequency measurements from 2005 July up to the last \xmm\
  observation ($\dot{\nu}=-1.35\times10^{-11}$~Hz~s$^{ -1}$). {\sl
    Bottom panel.} Blue triangles represent instantaneous frequency
  derivative between two consecutive frequency measurements (excluding
  \xmm\ data). Red triangles are the frequency derivative between two
  consecutive \xmm\ observations. Note the increase in $\dot{\nu}$
  after 2004 and the subsequent decrease around mid 2005. $\dot{\nu}$
  remained more or less constant until the last \xmm\ observation, with
  some variation around mid 2008.}
\label{freqHis}
\end{center}
\end{figure}

We first applied a barycenter correction to the filtered pn event
files. We then extracted a light curve (LC) for each of the 14
observations in the energy range 1.5$-$10~keV, at the 128-ms and 64-ms
resolution for large window and small window mode observations,
respectively. We ran the SAS task {\sl epiclccorr} on these LCs to
correct their count rates for background and for the events lost due
to various mirror
inefficiencies\footnote{http://xmm.esac.esa.int/sas/current/doc/epiclccorr/}. We
performed epoch-folding to derive an initial spin period for each of
the observations. Then, we split each observation into 5 smaller
intervals, and phase-connected those to achieve a better spin-period
measurement and smaller error. The results of this analysis are
summarized in Table~\ref{logObs}.

\begin{figure*}[ht]
\begin{center}
\includegraphics[angle=0,width=0.95\textwidth]{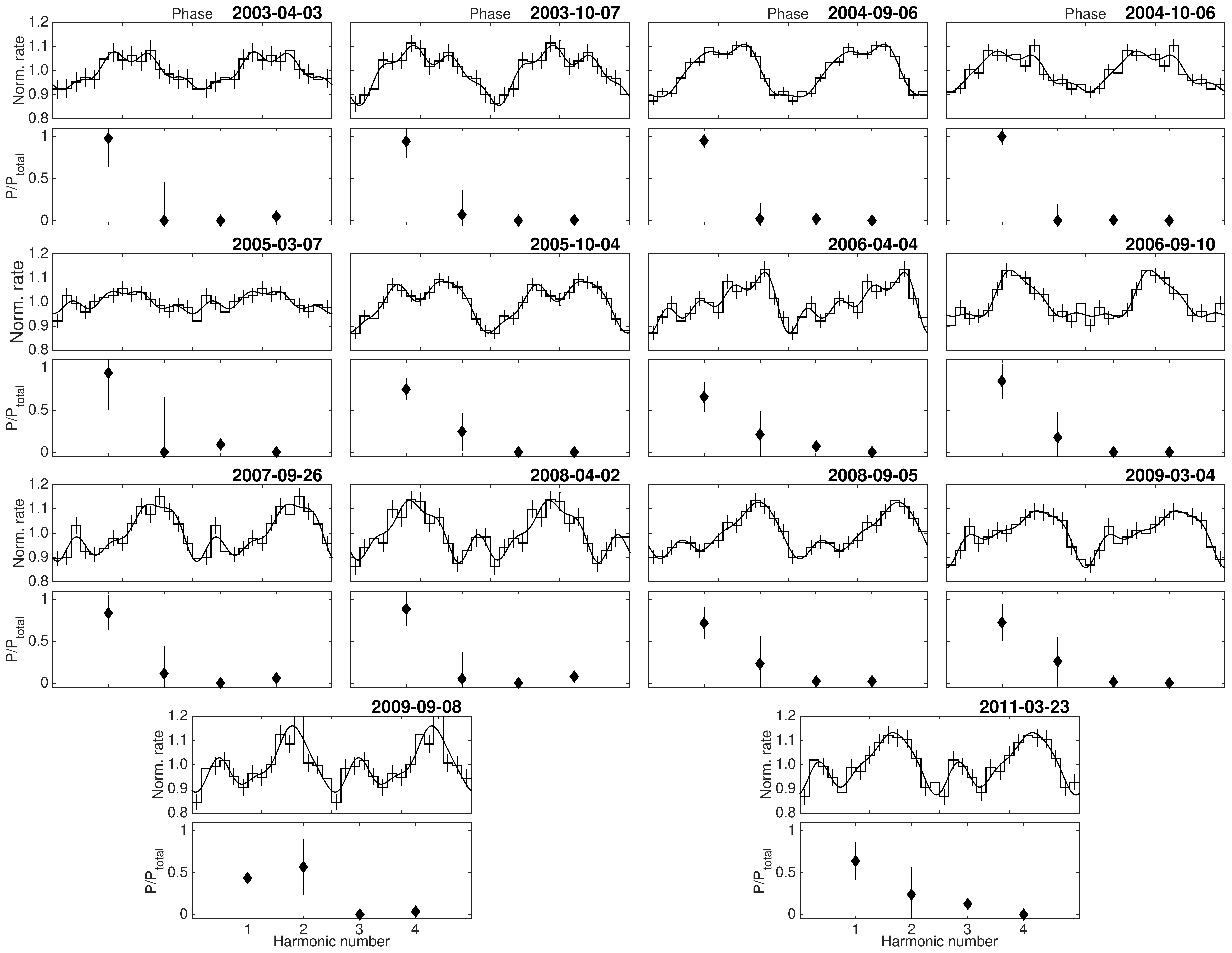}
\caption{The background subtracted PPs of all \xmm\ observations in
  the 1.5-10~keV, along with their different harmonic
  contributions. Two cycles are shown for clarity. The  pre-GF (prior
  to 2005) PP has a single broad-peak morphology. It becomes more
  complex afterwards with contribution from at least the second
  harmonic. For clarity purposes, we omitted the error bars on the
  power of harmonics 3 and 4 due to their large uncertainties.}
\label{PPALL}
\end{center}
\end{figure*}

We used our data to extend the work of \citet{woods07ApJ:1806}. In
that paper, the authors built a comprehensive frequency and frequency
derivative history of \src\ from late 1993 up to early 2005, a few
months after the GF. In Figure~\ref{freqHis}, we show the timing
history of \src\ extending to the last \xmm\ observation (red dots),
almost six and a half years after the GF. The blue triangles in the
bottom panel of Figure~\ref{freqHis} show the instantaneous frequency
derivative calculated between two adjacent frequency data points,
excluding the \xmm\ data. The red triangles show the frequency
derivative as derived from consecutive \xmm\ observations\footnote{It
  is obvious here that the early \xmm\ observations would have
  completely missed the timing noise of the source observed with \rxte\
  in 2004. We note, though, that the \xmm\ data are in agreement with
  the surrounding \rxte\ points.}.

The combined \rxte\ and \xmm\ data show some interesting trends; the
frequency derivative, almost one year after the GF decreased to the
minimum pre-flare value (i.e., maximum spin-down rate; note the
negative y-axis values), and remained at that level up to the last
\xmm\ observation. The dashed and dotted lines in Figure~\ref{freqHis}
are fits to the frequency measurements from 1993 to 2000 January,
$\dot{\nu}=-1.48\times10^{-12}$~Hz~s$^{ -1}$, and 2001 January to 2004
April, $\dot{\nu}=-8.69\times10^{-12}$~Hz~s$^{ -1}$, respectively
\citep{woods07ApJ:1806}. The dot-dashed line is our best fit to
frequency measurements from 2005 July up to the last \xmm\
observation, $\dot{\nu}=-1.35\times10^{-11}$~Hz~s$^{ -1}$. This latter
$\dot{\nu}$ value is 1.6 times larger than the one observed between
2000 and 2004, and almost an order of magnitude larger than the
historical level observed prior to 2000. Deviation from the average
after 2005 is seen around 2008 when $\dot{\nu}$ decreased to its
lowest value over the course of 8 years (i.e., spin-down rate attained
a maximum level), but recovered during the subsequent
observation. Unfortunately, the sparse \xmm\ observations around that
time did not allow for detailed modeling of this variation. It is
interesting, however, that it happened following a modest bursting
episode, as seems to be the case of the gradual change observed in
1999.

\begin{figure*}[ht]
\begin{center}
\includegraphics[angle=0,width=0.95\textwidth]{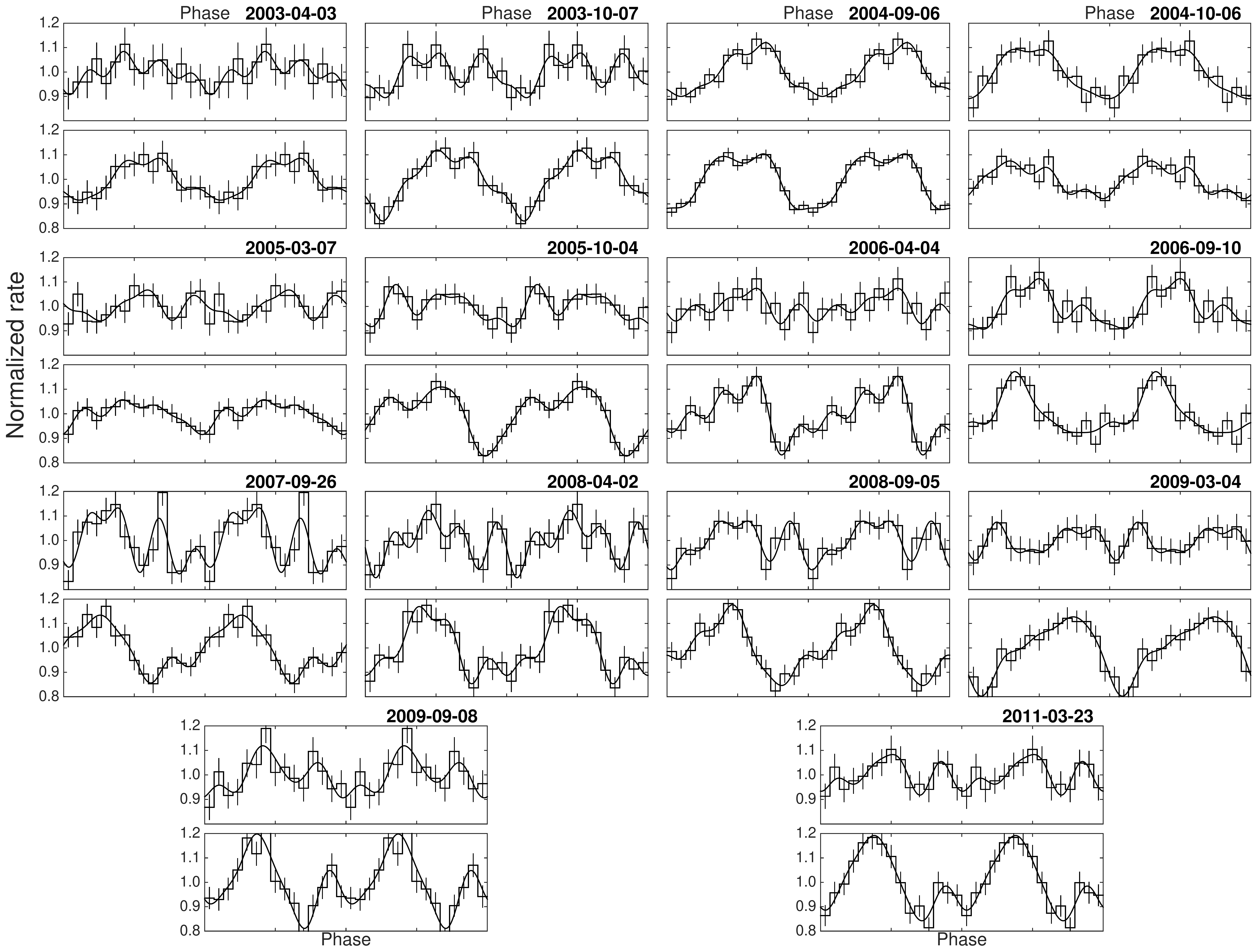}
\caption{The background-subtracted PP of all \xmm\ observations in two
  energy bands. Two cycles are shown for clarity. The upper rows show
  the 1.5-4~keV range for each observation and the lower rows show the
  4-10~keV range. See text for details.}
\label{PPALL_diffBands}
\end{center}
\end{figure*}

For completeness, we added in the top panel of Figure~\ref{freqHis}
the number of bursts detected in each of the \xmm\ observations (red
bars) and the bursts reported in GCNs (Gamma-ray Coordinates Network)
from 2005 up to April 2011 (these are mostly bursts identified with
the InterPlanetary Network\footnote{We caution that this list of
  bursts, e.g., Figure~\ref{freqHis}, could be missing a few that were
  not reported in GCNs, however it is a good representation of the
  frequency of the source bursting activity at a given time.}).

Using the spin periods of the different observations, we epoch-folded
the data to compute the pulse profiles (PPs) in the 1.5$-$10~keV
energy range (Figure~\ref{PPALL}). We fit the different PPs with a
sine plus cosine function \citep[e.g., ][]{
  bildsten97ApJ:PP,younes15ApJ:1744}, including contribution from four
harmonics. In Figure~\ref{PPALL}, we show the power of the different
Fourier components relative to the total power in the signal. Similar
to the findings of \citet[][see also
\citealt{mereghetti05ApJ:1806,woods07ApJ:1806}]{gogus02ApJ:1806}, we
find that the PP prior to the December 2004 GF has a single broad-peak
shape with negligible contribution from higher harmonics
(Figure~\ref{PPALL}). During and after the sixth \xmm\  observation,
i.e., 10 months after the GF, the PP became more complex, with
contribution from the second harmonic, indicating a multi-peak
structure, clearly seen in the PP (Figure~\ref{PPALL}).

We then derived the rms pulsed fraction (PF, e.g., \citealt{
  bildsten97ApJ:PP}) of the different observations in the 1.5$-$10~keV
energy range, and we estimated its error by simulating 1000
PPs. Despite the changes in PP morphology over the 8 years, we find a
steady PF of $\sim$7\%. This is consistent with the historical level
measured with \asca\ and \sax\ in 1995. The only change in PF was
observed immediately after the GF, when it dropped to a minimum of
3\%\ \citep[observation ID 0164561301, 2005-03-07, see also ][]{
  rea05ApJ:1806}.

Finally, we looked at the PP morphology in two different energy bands,
1.5$-$4~keV and 4$-$10~keV (Figure~\ref{PPALL_diffBands}). The two
observations of September and October 2004, when the source flux was
at a maximum, showed somewhat similar shapes and PFs in both bands,
i.e., a single broad peak. In all other observations, the PPs differed
between the two bands (the soft PP was more complex in general than
the hard one), and the soft band PF was lower than that of the hard
band. However, we note here that due to the low statistics of the
data, the soft band rms PFs were not very well constrained and their
$3\sigma$ upper limits are consistent with the hard band PFs.

\subsection{Spectral analysis}
\label{specana}

\begin{figure}[!th]
\begin{center}
\includegraphics[angle=0,width=0.49\textwidth]{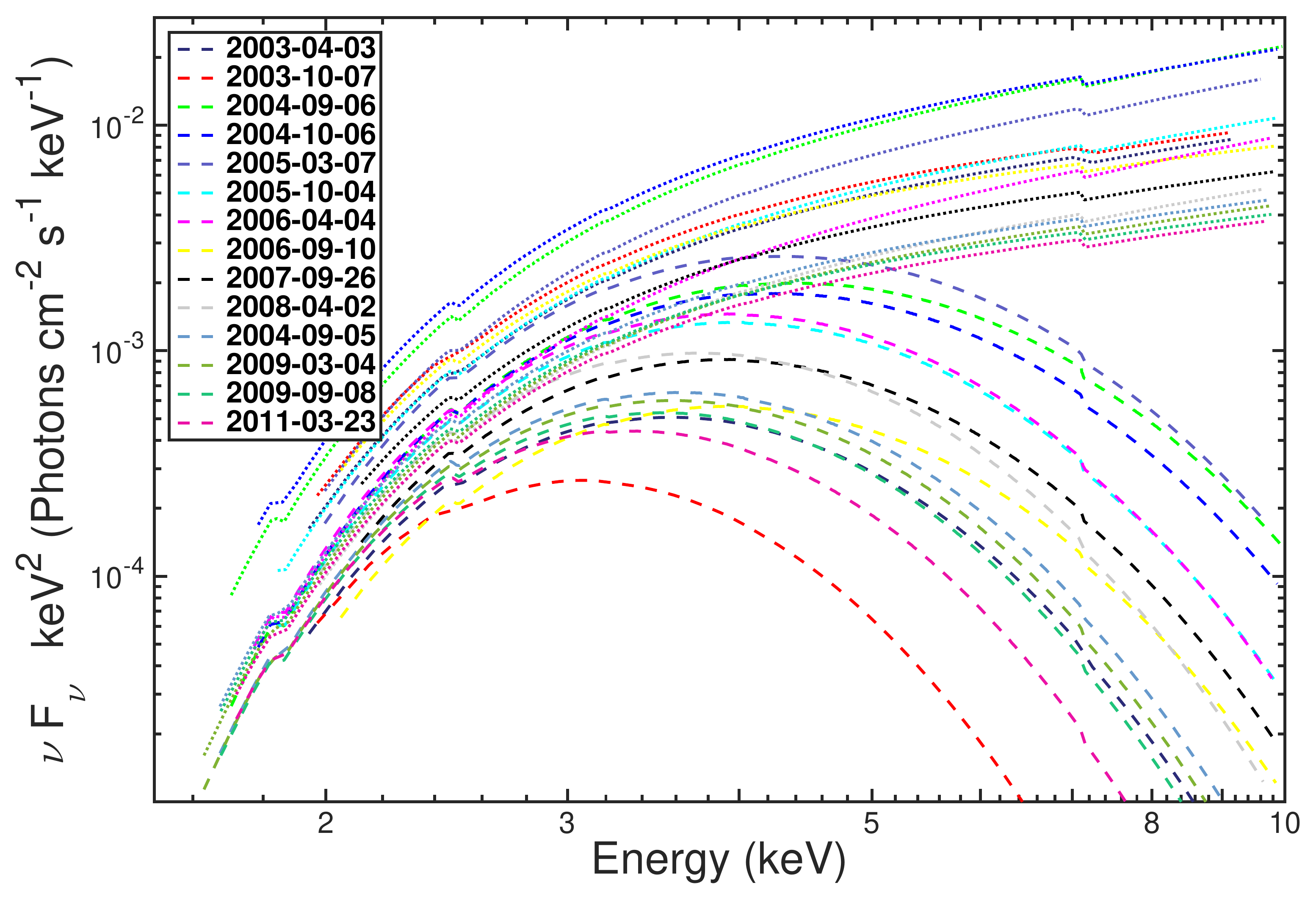}
\caption{$\nu F\nu$ best fit model of all \xmm\ observations. The
  separate BB and PL components are shown in dashed and dotted lines,
  respectively. Notice the spectral evolution of the BB component with
  flux, i.e., the cooling trend with decreasing flux. See text for
  more details.}
\label{specFit}
\end{center}
\end{figure}

We fit the pn spectra of all 14 observations simultaneously. We
started with a simple absorbed PL model and left the PL indices and
normalizations free to vary, while we linked the absorption between
all spectra. We find an acceptable fit with a $\chi^2$ of 5402 for
5284 degrees of freedom (d.o.f.). The hydrogen column density is
$N_{\rm H}=(9.7\pm0.1)\times10^{22}$~cm$^{-2}$. This simple model,
however, resulted in strong residuals in the form of a wiggle in the
spectrum. We then added a BB component to the model
\citep[e.g.,][]{mereghetti05ApJ:1806}. This more complicated spectral
shape provided a better fit to the data with a  $\chi^2$ of 5180 for
5256 d.o.f. According to an F-test, the improvement of the PL+BB model
over the single PL has a chance occurrence of $5.3\times10^{-32}$,
hence, we conclude that the addition of the BB is highly
significant. The spectra in $\nu{\rm F}_\nu$ space, are shown in
Figure~\ref{specFit}, while the spectral fit results are summarized in
Table~\ref{specParam}.

\begin{table}[!t]
\caption{BB and PL spectral parameters}
\label{specParam}
\newcommand\T{\rule{0pt}{2.6ex}}
\newcommand\B{\rule[-1.2ex]{0pt}{0pt}}
\begin{center}{
\resizebox{0.49\textwidth}{!}{
\begin{tabular}{c c c c c c}
\hline
\hline
Observation ID   \T\B & kT     & R$^{2,a}$       & $\log F_{\rm BB}$ & $\Gamma$ & $\log F_{\rm PL}$ \\
                           \T\B & (keV) & (km$^2$) & (erg s$^{-1}$ cm$^{-2}$) & & (erg s$^{-1}$ cm$^{-2}$)\\
\hline
0148210101 \T & $0.6_{-0.2}^{+0.1}$     & $1.2_{-0.9}^{+1.4}$ & $-11.6_{-0.7}^{+0.3}$ & $1.3\pm0.2$ & $-10.70_{-0.07}^{+0.06}$\\
0148210401 \T & $0.5\pm0.2$     & $3.2_{-2.5}^{+5.7}$ &$-11.8\pm0.6$ & $1.4_{-0.2}^{+0.1}$ & $-10.63_{-0.06}^{+0.04}$\\
0205350101 \T & $0.90\pm0.05$ & $0.7\pm0.2$ & $-11.23\pm0.02$ &$1.1\pm0.1$ & $-10.41\pm0.02$\\
0164561101 \T & $0.85\pm0.06$ & $0.9\pm0.3$ & $-11.28\pm0.03$ &$1.2\pm0.1$ & $-10.37\pm0.02$ \\
0164561301 \T & $0.86\pm0.06$ & $1.1_{-0.5}^{+0.4}$ & $-11.1\pm0.2$  & $1.0_{-0.3}^{+0.2}$ & $-10.55\pm0.08$ \\
0164561401 \T & $0.77\pm0.04$ & $1.1\pm0.3$& $-11.34\pm0.02$&$1.2\pm0.1$& $-10.68\pm0.03$ \\
0406600301 \T & $0.75\pm0.03$ & $1.2\pm0.2$ &$-11.28\pm0.02$& $1.0_{-0.2}^{+0.1}$&$-10.83_{-0.04}^{+0.03}$\\
0406600401 \T & $0.75\pm0.08$ & $0.5\pm0.3$ & $-11.29\pm0.02$&$1.5\pm0.1$&$-10.67\pm0.03$ \\
0502170301 \T & $0.74\pm0.04$ & $0.9\pm0.3$ & $-11.48\pm0.03$ &$1.4_{-0.2}^{+0.1}$ & $-10.83\pm0.04$ \\
0502170401 \T & $0.69_{-0.04}^{+0.03}$ & $1.4\pm0.3$ & $-11.45\pm0.02$ &$1.2\pm0.2$ & $-10.98\pm0.05$ \\
0554600301 \T & $0.65_{-0.04}^{+0.03}$ & $1.3\pm0.3$ & $-11.58\pm0.03$ &$1.4\pm0.1$ & $-10.93_{-0.04}^{+0.03}$ \\
0554600401 \T & $0.63\pm0.04$ & $1.3_{-0.3}^{+0.4}$ & $-11.60\pm0.03$ &$1.4\pm0.1$ & $-10.99\pm0.04$ \\
0604090201 \T & $0.61\pm0.04$ & $1.4_{-0.4}^{+0.5}$ &$-11.65\pm0.03$ & $1.5\pm0.1$ & $-10.98\pm0.04$ \\
0654230401 \T & $0.56\pm0.04$ & $2.0_{-0.5}^{+0.7}$ & $-11.64_{-0.03}^{+0.04}$&$1.4\pm0.1$ & $-11.03\pm0.04$ \\
\hline
\end{tabular}}}
\end{center}
\begin{list}{}{}
\item[{\bf Notes.}]$^{a}$Derived by adopting an 8.7~kpc distance 
  \citep{bibby08MNRAS:1806}. Fluxes are in the energy range 0.5-10~keV.
\end{list}
\end{table}

\begin{figure*}[h]
\begin{center}
\includegraphics[angle=0,width=0.97\textwidth]{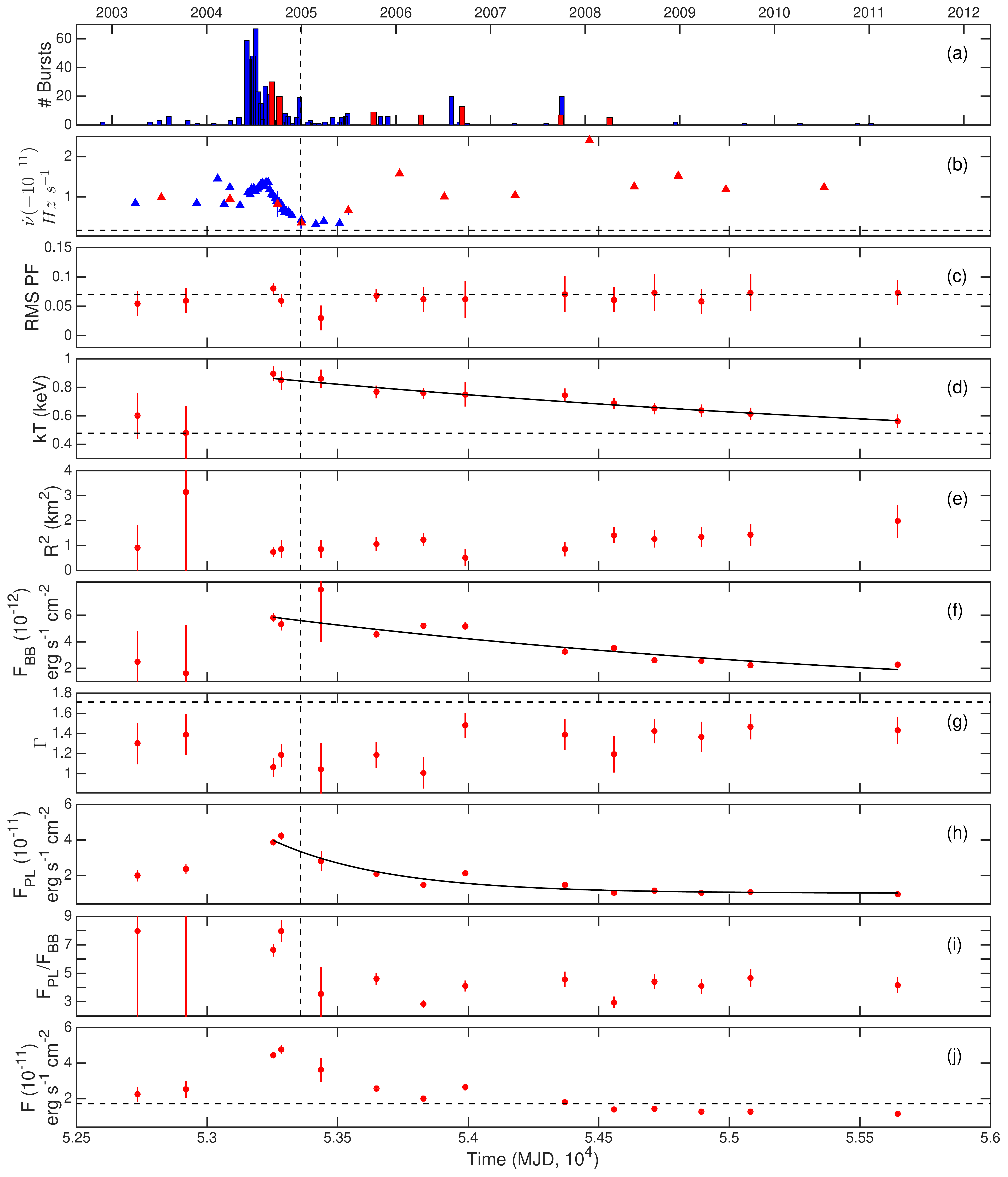}
\caption{Temporal and spectral histories of \src. {\sl Panel (a).}
  Number of bursts in 30 days as seen from wide field-of-view
  instruments (blue bars), and bursts detected by \xmm\ (red)
  bars. {\sl Panel (b).} Frequency derivative from \citet{
    woods07ApJ:1806} (blue triangles), and from \xmm\ (red
  triangles). {\sl Panel (c).} RMS PF. {\sl Panel (d).} BB
  temperature, $kT$. {\sl Panel (e).} BB area, $R^2$. {\sl Panel (f).} BB
  flux, $F_{\rm BB}$. {\sl Panel (g).} PL index, $\Gamma$. {\sl Panel (h).} PL
  flux, $F_{\rm PL}$. {\sl Panel (i).} PL to BB flux ratio. {\sl Panel
    (j).} Total flux. In all panels, a dotted horizontal line
  represents the historical level of the given parameter measured with
  \asca\ and/or \sax. The vertical dashed line is the GF epoch. Solid
  lines are exponential decay trends for given parameters following
  the 2004 outburst. See text for details.}
\label{timeSpecProp}
\end{center}
\end{figure*}

Figure~\ref{specFit} shows a clear spectral evolution between the
different \xmm\ observations, in both the PL and the BB components. We
plotted the evolution of the different spectral (and temporal)
parameters in  Figure~\ref{timeSpecProp}. The BB and PL fluxes
increased by a factor of $\sim$2.5 and 4, respectively, compared to
their quiescent level, following the onset of the 2004 bursting
activity from the source (see  {\sl panel (a)} and, e.g., \citealt{
  mereghetti05ApJ:1806}). This flux increase is clearly accompanied by
spectral hardening in both the BB and the PL components
(Figure~\ref{timeSpecProp}).

The spectral parameters decayed back to quiescence levels
quasi-exponentially. We fit the PL and the BB fluxes with a model of
the form $F(t)\propto \exp^{-t/\tau}$ (black solid lines, {\sl panels
  (f)} and {\sl (h)}, Figure~\ref{timeSpecProp}), where $\tau$ is the
characteristic decay time-scale representing a 63\%\ decay of a given
parameter back to quiescence. For the PL flux, we find a
characteristic decay time-scale of $\tau=441_{-91}^{+104}$~days. The
characteristic decay time-scale of the BB flux was much less 
constrained, with $\tau=990_{-533}^{+509}$~days. We find a similar
decay time-scale for the BB temperature, $kT$ ({\sl panel (d)},
Figure~\ref{timeSpecProp}). A power-law (PL) decay trend, $F(t)\propto
t^{\alpha}$ for the above parameters results in a PL decay index
$\alpha=-0.65\pm0.1$ for the PL flux and $-0.4\pm0.2$ for the BB
flux. Finally, we derive the total energy emitted in each component
since MJD 53126 (2004 May 01, i.e., the onset of the major bursting
activity of 2004) assuming an exponential decay trend, and we find
$E_{\rm tot,~PL}=3.3\times10^{43}$~ergs and $E_{\rm
  tot,~BB}=0.7\times10^{43}$~ergs for the PL and BB components,
respectively. We find similar values when using the PL decay trend.

It is worth noting that the BB area increased during the outburst in
tandem with the flux decrease from $R^2\approx0.7$~km$^2$ to
$R^2\approx2.0$~km$^2$. Such a behavior has been seen before in a few
magnetars \citep[e.g.,
][]{woods04ApJ:1E2259,israel10MNRAS:1E1547}. Moreover, the PL to BB
flux ratio reached maximum during the beginning of the outburst
($(F_{\rm PL}/F_{\rm BB})\approx7$), where the PL flux increased by a
factor 1.6 more than the BB flux ({\sl panel (i)},
Figure~\ref{timeSpecProp}). The ratio remained constant after the 2006
observation ($(F_{\rm PL}/F_{\rm BB})\approx4.5$).

Finally, we performed phase-resolved spectroscopy on all
observations. We split each observation into 10 equally separated bins
in phase space and fit them simultaneously. We linked the absorbing
column between the different spectra, and let the BB and PL spectral
parameters free to vary. The only clear trend we observe is in the
2004 September and October observations. We find that only the PL flux
shows any phase variability \citep{mereghetti05ApJ:1806}, possibly
indicating that the PL component is responsible for most of the pulsed
flux. In the remaining cases, unfortunately, the combination of worse
statistics compared to the 2004 observations and the very low PF of
the source hindered any meaningful conclusions.

\section{Discussion}
\label{discuss}

\src\ is one of the most active magnetars, exhibiting continuous
bursting activity, either as strong outbursts like the one detected in
2004, or in sporadic groups of few isolated bursts
(Figures~\ref{freqHis} and \ref{timeSpecProp}, upper panels). \src\
also shows one of the most erratic behaviors in its timing properties
\citep[Figure~\ref{freqHis}, ][]{woods07ApJ:1806}. The very early
observations of the source
\citep{kouveliotou98Nat:1806,woods00ApJ:1806,mereghetti00AA:1806}
between 1993 and 1999 showed a steady spin-down,
$\dot{\nu}=-1.48\times10^{-12}$~Hz~s$^{-1}$. From early 1999 to late
2000, the spin-down rate slowly increased to a new steady state,
$\dot{\nu}=-8.69\times10^{-12}$~Hz~s$^{-1}$, that persisted up to the
outburst emitted by the source starting roughly in mid
2004. Thereafter, the source exhibited an erratic behavior leading to
another dramatic change in the $\dot{\nu}$ trend
\citep{woods07ApJ:1806}. Our follow-up analysis shows that the
spin-down of the source increased again to new long-term average
between July 2005 and March 2011 with a new value of
$\dot{\nu}=-1.35\times10^{-11}$~Hz~s$^{-1}$.

The current \src\ spin-down rate is an order of magnitude larger than
its first historical value. Assuming that the spin-down rate is due to
magnetic dipole radiation, the average surface dipole magnetic field
strength of a magnetar is
$B\approx3.2\times10^{19}(\dot{\nu}/\nu^3)^{1/2}$, where $\nu=1/P$ and
$P$ is the spin period of the source. An increase, therefore, of 
$\dot{\nu}$ by a factor of 10 in a span of 5.5 years would imply an
increase of 3.2 in dipole magnetic field strength, which is rather
unlikely. We explore below possible mechanisms that may have
contributed to the torque increase on the star, hence increasing its
spin-down rate.

\citet[][also see
\citealt{harding99ApJ:mag,thompson98PhRvD:mag}]{thompson00ApJ:1900} 
suggested that persistent luminosity of Alfven waves and particles
could generate an additional torque (besides the magnetic dipole one)
that would affect the spin-down rate of a magnetar. In this model,
continuous low-level seismic activity in a slowly spinning
highly-magnetized neutron star will excite magnetospheric Alfven modes
and induce a relativistic particle outflow out to a large radius. This
particle outflow will import extra torque on the star, increasing its
spin-down rate. According to \citet{thompson00ApJ:1900}, the spin-down
torque increases by a factor of $\sim(L_{\rm A}/L_{\rm MDR})^{1/2}$,
where $L_{\rm A}$ is the persistent luminosity of Alfven waves and
particles, and $L_{\rm MDR}$ is the magnetic dipole luminosity (the
time-average X-ray output of the SGR in quiescence derived using its
dipole field $B$ and spin period, \citealt{thompson96ApJ:magnetar}). An
order of magnitude increase in spin-down from 2000 to $\sim$mid 2005
would require an increase in the particle luminosity by a factor of
100 over $L_{\rm MDR}$ (considering the latter to correspond to the
2000 early X-ray observations). An obvious effect of such a strong
particle output would be a luminous wind nebula \citep[][see also
\citealt{younes12ApJ:1834}]{thompson00ApJ:1900,tong13ApJ:wind}. At the
8.7~kpc distance of \src, the angular extent of such a wind nebula
would be of the order of tens of arcseconds when considering
relativistic speeds for the particle outflow, which could possibly be
observed with \chandra. Such a wind nebula is, however, not detected
in \src\ \citep{vigano14:1806}.

A twisted magnetosphere could also impart excess torque on a
magnetar. Magnetic stresses inside the star will cause motion to the
footprints of the surface magnetic field lines, causing these external
fields to twist \citep{thompson02ApJ:magnetars,beloborodov09ApJ}. Such
a twisted configuration causes a slower decrease in the dipole
magnetic field with increasing distance from the neutron star
(compared to the case where no twist exists), resulting in a stronger
$B$ value at the light cylinder. This excess magnetic energy is then
dissipated through an increase in the spin-down rate
\citep{thompson02ApJ:magnetars}. However, if the affected twisted
fields are closed field lines, this temporal variation should also be
accompanied by variation in the persistent X-ray spectrum, since a
twisted closed field line will have an increase in the optical depth
to resonant scattering \citep{thompson02ApJ:magnetars,
  beloborodov09ApJ}. On the other hand, if the magnetic flux is dissipated
through open field lines, it would affect the torques on the star at
large radii without having any effect on the persistent X-ray spectrum
\citep[e.g., SGR~J1745$-$2900, ][]{kaspi14ApJ:1745}. The spin-down
should return to its pure dipole value as the twist angle
decreases. The time-scale for the twist energy dissipation could be
shorter than the spin period of the source, if the twist is large
enough that most of the energy is dissipated through a large magnetic
reconnection event, or, more interestingly, for small twist angles,
the twist decay time-scale could be as long as a few years
\citep{parfrey13ApJ,parfrey12ApJ,beloborodov07ApJ:magCorona}. This
model could explain the continuous large spin-down rate of \src\ up to
mid 2011.

We now discuss the long-term spectral evolution of the \src\ X-ray
emission after its 2004 major bursting episode\footnote{The short-term
  evolution of \src\ spectral properties after the 2004 outburst were
  published in \citet{woods07ApJ:1806}.}.  \src, similar to many other
magnetar sources, has a 1-10~keV spectrum well fit with the
combination of a thermal (BB), and a non-thermal (PL) component. In
the magnetar model, the BB component is due to internal heating from
the decay of the strong magnetic field, while the non-thermal
component is due to resonant cyclotron scattering of these thermal
photons by the plasma in the magnetosphere
\citep[e.g.,][]{thompson02ApJ:magnetars}. After the 2004 bursting
episode, the flux of the BB and PL increased, while showing spectral
hardening (BB temperature increased and PL index, $\Gamma$,
decreased). Such a flux-hardness relation is a prediction of the model
presented in \citet{thompson02ApJ:magnetars} and a very common
phenomenon among magnetar sources \citep[e.g., ][]{gavriil04ApJ:1048,
  scholz11ApJ:1547}.

One of the models usually invoked to explain the outburst evolution of
a magnetar is of external magnetospheric origin
\citep{beloborodov09ApJ}. In this model, a twist in the magnetosphere,
if acting on a bundle of closed field lines, would increase their
particle charge density. Returning currents from these field lines
would hit their surface footprints, covering a certain area $A$ on the
surface, heating it and causing radiation of thermal photons. These
thermal photons, in turn, are Compton scattered to higher energies by
the plasma in the bundle. As the twist relaxes back to its original
untwisted configuration, the bundle disappears gradually, and both the
thermal and non-thermal components decrease back to their quiescent
level. The relaxation time-scale of the bundle,
$t\approx10^{7}\mu_{32}\Phi_{10}^{-1}A_{11.5}$~s, depends primarily on
the footprint size $A_{11.5}=(A/10^{11.5})$~cm$^2$, the magnetic
dipole moment $\mu_{32}=(\mu/10^{32})$~G~cm$^{3}$, and the electric
voltage sustaining its plasma density, $\Phi_{10}=(\Phi/10^{10})$~V
\citep{beloborodov09ApJ,mori13ApJ:1745}. Considering a large dipole
field $B=10^{15}$~G, a voltage $\Phi=10^{9}$~V, and the area of the BB
component $A=4\pi R^2$ with $R^2\approx1$~km$^2$
(Table~\ref{specParam}), one can roughly reconcile the long relaxation
time for \src, $\sim10^8$~s, with the one predicted by this
model. However, the model also predicts that $A$ decreases with
decreasing flux, which we do not observe in the data
(Figure~\ref{timeSpecProp}). Moreover, the PL flux increased by a
factor of $\sim1.6$ more than the BB after the onset of the outburst
(the 2004 observations), compared to the following observations (after
October 2005) where their ratio was more or less constant - a behavior
that does not seem to agree with the twisted magnetospheric model.

Heating of the crust from re-arragement of the internal magnetic
field, e.g., due to a sudden crack of the crust, is also used to
reproduce the relaxation light curve of magnetars \citep[e.g.,
][]{lyubarski02ApJ:coolMag,kouveliotou03ApJ:1627,pons12ApJ:mag}. In
this picture, a total energy of the order of $\sim10^{43}$~ergs is
suddenly deposited into the crust, and re-radiated gradually in the
form of thermal photons. The total energy emitted from \src\
throughout the outburst supports these numbers. However, the decay
time-scale for such models is expected to be much shorter (of the
order of $\sim100$~days) than the years time-scale we calculate for
\src\ \citep[see discussion in ][]{zelati15MNRAS:1745}.

It is remarkable that the PF of \src\ has remained at a constant value
of about 7\%\ from 1994 to 2011 (it only changed for a short period of
time immediately after the GF, \citealt{rea05ApJ:1806}), while all its
other properties changed drastically. This could be an indication that
the geometry (and location, e.g., close to the magnetic poles) of the
emitting region has remained essentially the same for the last two
decades.

Finally, we note that there has been attempts to link the intrinsic
properties of magnetars (e.g., $\dot{\nu}$, $B$) with their observed
X-ray properties \citep{marsden01ApJ,kaspi10ApJ,an12ApJ:1627}. Such
correlations assume that the measured intrinsic properties of these
sources outside of bursting episodes are their ``true'' values
(assuming a dipole configuration). It is clear from our analysis of
\src\ (see also \citealt{woods07ApJ:1806}) that such assumption is not 
necessarily correct. The spin-down rate we measure for the late \xmm\
observations - clearly outside of major bursting episodes - represents
an order of magnitude variation in about six years. Hence, the
``true'' $\dot{\nu}$, and by extrapolation $B$, are currently unknown
for \src. This could also be true for other magnetars. The above
mentioned correlations need to take such possibilities into account.

\section*{Acknowledgments}

This work is based on observations with \xmm - principal investigator,
Sandro Mereghetti - an ESA science mission with instruments and
contributions directly funded by ESA Member States and the USA
(NASA). V. M. K. acknowledges support from an NSERC Discovery Grant
and Accelerator Supplement, the FQRNT Centre de Recherche
Astrophysique du Qu\'ebec,  an R. Howard Webster Foundation Fellowship
from the Canadian Institute for Advanced Research (CIFAR), the Canada
Research Chairs Program and the Lorne Trottier Chair in Astrophysics
and Cosmology. We thank the referee for a careful read of the
manuscript and their constructive comments that improved the quality
of the article.

\end{document}